\documentstyle[preprint,aps]{revtex}

\begin{document}

\preprint{Usach/95/08}

\title{ON THE INITIAL SINGULARITY PROBLEM IN \\ 
TWO-DIMENSIONAL QUANTUM COSMOLOGY}

\author{J. GAMBOA\thanks{E-mail: jgamboa@lauca.usach.cl }}

\address{Departamento de F\'{\i}sica, Universidad de Santiago 
de Chile\\ 
Casilla 307, Santiago, Chile }

\maketitle

\begin{abstract} 

The problem of how to put interactions in two-dimensional
quantum gravity in the strong coupling regime is studied. It
shows that the most general interaction consistent with this 
symmetry is a Liouville term that contain two
parameters $(\alpha, \beta)$ satisfying the algebraic relation 
$2\beta - \alpha =2$ in order to assure the closure of the 
diffeomorphism algebra. 

The model is classically soluble and it contains as 
general solution the temporal singularity. The theory is
quantized and we show that the propagation amplitude fall to
zero in $\tau =0$. This result shows that the classical
singularities are smoothed by quantum effects and the bing-bang 
concept could be considered as a classical extrapolation instead
of a physical concept.  
\end{abstract}

The quantization of the gravitational field is a problem that has resisted a 
solution for many years is spite of the intense research in this field 
\cite{alvarez}.

The black holes evaporation \cite{hawking1}, loss of information 
\cite{hawking2} or the quantum non-hair theorems \cite{coleman} are 
ideas that emerged in the last two decades for which there are not still 
a definitive explanation in terms of a true quantum theory of gravity.

However in spite of these results are very promisories,
fundamentals problems associated to the physical interpretation
of the theory are very far to be reached and it seems to
indicate that is necessary to introduce significatives
simplifications in order to get a complete understanding of the
theory. 

In \cite{yo} we proposed a two-dimensional quantum gravity model
in the strong coupling regime (SCR) which was mapped to a null
string theory embeded in a two-dimensional target space. In this
model, the temporal component of the bi-vector on the target
$\Phi_A$ was identified as the time in quantum gravity and, as a
consequence, one could define an evolution operator or a
probability concept as in ordinary quantum mechanics.

The purpose of the present letter is to extend our previous
analysis in order to include interactions in the SCR respecting 
all the symmetries of the model. In particular, we will show
below that the most general interaction consistent with the
reparametrization invariance is a Liouville-like one, this
interaction is very interesting because will permit to make
contact with classical and quantum cosmology.

More precisely, we will show that the general solutions of the
gravitational field equations  are singular in $\tau = 0$, {\it 
i.e.} in two dimensions there is a Kasner-like solution although
it has not an oscillatory behaviour as in four dimensions. The
theory also can be exactly quantized and the propagation
amplitude reflects the fact at quantum level, the classical 
singularities disappears\footnote{Here one could ask if this
result remain in other 2D gravity models. The answer is yes
although the demostration must be performed model by model (an
exception is the Jackiw model)}.

In the strong coupling limit $2D$ gravity is described by the 
constraints 

\begin{eqnarray}
{\cal H}_{\perp} &=& 
{1\over 2} \biggl[ b^2 -4P^2 \biggr],
\nonumber \\  
{{\cal H}}_1 &=& b \psi^{'} + P {\cal H}i^{'} - 2P^{'} + b^{'}, \label{a10}
\end{eqnarray}
where $(b, \psi)$ and $(P, {\cal H}i)$ are canonical variables and
the constraints (\ref{a10}) satisfy the diffeomorphism algebra 
\footnote{For conventions and notation see
\cite{yo}, however we will emphasize that the fields $\psi$ and
${\cal H}i$ (in the proper time gauge are related with $g_{\mu \nu}$
by means  
$ g_{00} =1, g_{01}=g_{10} = 0, g_{11} = e^\chi$.}
 
\begin{eqnarray} 
& \lbrack& {\cal H}_{\perp} (x), {\cal H}_{\perp} (x^{'}) \rbrack = 0,
\nonumber 
\\  
&\lbrack& {\cal H}_{\perp} (x), {\cal H}_1 (x^{'}) \rbrack = 
( {\cal H}_{\perp} (x)
+ {\cal H}_{\perp} (x^{'} )) \delta^{'} (x - x^{'}), \label{a11} \\ 
& \lbrack& {\cal H}_1 (x) , {\cal H}_1 (x^{'}) \rbrack =
( {\cal H}_1 (x) + {\cal H}_1 (x^{'})) \delta^{'} (x - x^{'}). \nonumber 
\end{eqnarray}

The next step is try to put interactions. One can use as a guide
principle that the possibles interactions in the model should 
respect the diffeomorphism symmetry, {\it i.e.} should preserve
the algebra (\ref{a11}). This fact, say us that the most general
constraints including interactions have the form    
\begin{eqnarray}
 {\cal H}_{\perp} &=& 
{1\over 2} [ b^2 -4P^2 ] + {1\over 8} \omega e^{{\alpha \psi + \beta \chi}},
\nonumber \\  
{{\cal H}}_1 &=& b \psi^{'} + P \chi^{'} - 2P^{'} + b^{'}, \label{a12}
\end{eqnarray}
where $\alpha, \beta$ and $\omega$ are constants. By a simple
dimensional analysis one can see that $\omega$ is a constant
that can be identified with the cosmological constant $\mu^2$,
whereas $\alpha$ and $\beta$ are two constants that
must be chosen satisfying the algebraic relation 
\begin{equation} 
2\beta - \alpha = 2, 
\end{equation} 
in order to preserve the closure of the diffeomorphism algebra.

This result is remarkable because it says that one can fix
appropiately $\alpha$ (or $\beta$) and to decouple the variables
of the theory.

Such is discussed in \cite{yo} the action   
\begin{equation} 
S~=~ \int d^2x~ [b{\dot \psi} + P{\dot \chi} - 
N{\cal H}_{\perp} - N_1 {\cal H}_1 ], 
\label{action} 
\end{equation} 
is invariant under the following transformations
\begin{eqnarray}
& \delta & \psi = \epsilon b + \epsilon_1 \psi^{'} - 
\epsilon_1^{'}, \nonumber 
\\ 
&\delta & \chi = -4\epsilon P + {\epsilon}_1 \chi^{'} - 2 {\epsilon}_1^{'}, 
\nonumber
\\ 
&\delta & b = -\alpha \epsilon \mu^2 e^{{\alpha \psi + \beta \chi}} + {({\epsilon}_1 b)}^{'} , 
\nonumber 
\\ 
&\delta & P = -\beta \epsilon \mu^2 e^{{\alpha \psi + \beta \chi}} + {(\epsilon_1 P)}^{'} 
\label{transfor}, 
\\ 
&\delta & N = {\dot \epsilon } + N^{'} \epsilon_1 - 
N \epsilon^{'}_1 + N^{'}\epsilon - N_1 \epsilon^{'},\nonumber
\\ 
&\delta & N_1 = {\dot {\epsilon}_1} + N_1^{'}\epsilon_1 - 
N_1 \epsilon^{'}_1,\nonumber
\end{eqnarray}
provided that $\epsilon (\tau_2, x) = 0 = \epsilon (\tau_1, x)$.  

Using (\ref{action}), the equations of motion after to eliminate the 
canonical momenta are
\begin{eqnarray} 
&{\displaystyle {\partial\over \partial \tau}} 
\biggl[ {{{\dot \psi} - N_1 \psi^{'} + 
N_1^{'}}\over N} \biggr] - {\displaystyle {\partial\over \partial x}} 
\biggl[ 
{{N_1({\dot \psi} - N_1 \psi^{'} + N^{'}_1)}\over N} \biggr] + {1\over 8} 
\mu^2 N e^{{\alpha \psi + \beta \chi}} = 0, \nonumber \\ & 
{\displaystyle {\partial\over \partial \tau}}
\biggl[ {{{\dot \chi} - N_1 \chi^{'} -2 N_1^{'}}
\over N} \biggr] - {\displaystyle {\partial\over \partial x}} 
\bigg[ 
{N_1({\dot \psi} - N_1 \chi^{'} -2 N^{'})\over N} \biggr] - 
{1\over 2}\mu^2 N e^{{\alpha \psi + \beta \chi}} = 0. \label{a7} 
\end{eqnarray}

These equations are difficult to solve in an arbitrary gauge,
but is suggested 
by a simple observation, one can solve it easily in the proper-time gauge 
\cite{claudio} 
\begin{equation} 
{\dot N} = 0, \,\,\,\,\,\,\,\,\,\,\,\, N_1 = 0. 
\end{equation}

Using this gauge condition, (\ref{a7}) becomes

\begin{eqnarray}
&{\ddot \psi} + {1\over 8} \alpha \mu^2 N^2 e^{{\alpha \psi + 
\beta \chi}} = 0,\nonumber \\ 
&{\ddot \chi} - {1\over 2}\mu^2 \beta N^2 e^{{\alpha \psi + 
\beta \chi}} = 0. \label{a8}
\end{eqnarray}

It is interesting to note that similar equations appear also in 
gravity in four dimensions when 
the behaviour of the solutions of the Einstein field equations near 
of the temporal singularity is studied. In fact, the Einstein field 
equations for a homogeneous and anisotropic space are \cite{landau}
\begin{eqnarray} 
& {\ddot a} = -{1\over 2} e^{4a}, \label{eq1}
\\ & 
{\ddot b} = {\ddot c} = {1\over 2} e^{4a}, \label{eq2}
\end{eqnarray}
where $a, b$ and $c$ are three functions that characterize 
the spatial metric tensor. The equation (\ref{eq1}) coincides with 
the \lq \lq Einstein equations" (\ref{a8}), while (\ref{eq2}) 
has not analog in two dimensions. The set of equations 
(\ref{eq1}-\ref{eq2}) is responsible for the oscillatory regime 
near of the time singularity \cite{bkl}. 
In our case at hand, the oscillatory regime is not 
possible due that there is not equation (\ref{eq2}), however as 
we will see below even so a Kasner-like solution will be found.

The equations (\ref{a8}) can be written in the following way
\begin{equation} 
{\ddot \rho} - {1\over 2} \gamma^2 e^\rho = 0, \label{a9}
\end{equation}
where $\rho = {\alpha \psi + \beta \chi}$ and 
$\gamma^2 = {1\over 4} \mu^2 N^2 (4\beta^2 - \alpha^2)$. 
The general solution is 
\begin{equation}  
\tau ~=~ {1\over \gamma} \int {d\rho \over \sqrt{D^2 + e^\rho }},
\label{a20}
\end{equation}  
where $D^2 = {A\over \gamma^2}$ and $A$ 
is an integration constant.

In order to integrate (\ref{a20}) three cases must be distinguished; 

i) $D^2<0$, {\it i.e.} when $A>0, \,\,\gamma^2<0$ or $A<0,\,\,
\gamma^2 >0$ and (\ref{a20}) gives
\begin{equation}
\gamma \tau = c + {2\over D} tan^{-1} \biggl[ {\sqrt{D^2 + e^\rho}\over D} 
\biggr], \label{mo1}
\end{equation}

ii) $D^2 >0$, {\it i.e.} when $A>0,\,\, \gamma^2 >0$ or 
$A<0,\,\, \gamma^2<0$ and 
(\ref{a20}) becomes
\begin{equation} 
\gamma \tau = c + 
{1\over D} \ln \biggl[ {{\sqrt{D^2 + e^\rho} -D}\over {\sqrt{D^2 + e^\rho} 
+ D}} \biggr]
\end{equation}

iii) $D^2 = 0$, {\it i.e.} $A=0$ and (\ref{a20}) is 
\begin{equation} 
\tau = c - \gamma e^{-\rho/2}. 
\end{equation}
where $c$ is a constant.

The cases considered above gives a relation between $\psi$ and $\chi$ but 
the general solution of the equation of motion are obtained inserting this 
relation back in the equation of motion for $\psi$ and $\chi$. 

The general solutions found are
\begin{eqnarray} 
 \chi &=& -{\mu^2 N^2 \beta\over 2\gamma^2} \ln \cos {({\gamma
D\over 2}\tau + c)} -
{\mu^2 N^2 \beta D^2\over 4} \tau^2 + \omega \tau +
\sigma, \label{model 1} 
\\  
\chi &=& -{2 \mu^2 N^2 \beta \over  \gamma^2} 
\ln \sinh {({\gamma D\over 2}\tau + c)}  + \omega \tau
+ \sigma, \label{model2} 
\\ 
\chi &=& -{\mu^2 \beta N^2 \gamma^2\over 2} 
\ln {(c - \tau )}  + \omega
\tau + \sigma, \label{model 3}
\end{eqnarray} 
for the models $i)-iii)$ and similar solutions for $\psi$ 
($\omega, \sigma$ are constants). 

From these results one see that there are elections for $\alpha$ (or
$\beta$) that gives singular solutions for $\tau =0$, in fact 
choosing
\begin{equation} 
{\beta\over {2\beta + \alpha}} >0, \,\,\,\,\,\,\,\,\,\,\,\, 
\beta (2\beta + \alpha) > 0, \label{desigual} 
\end{equation} 
and $\omega =0=\sigma$ for the models $ii)$ and $iii)$, one see
that the general solution 
near of $\tau =0$ in a synchronous frame has the form 
\footnote{It is 
remarkable to note that this solution corresponds to a 
Poincar\'e metric when $p=2$ after to make the transformation 
$t = \sqrt{2 \tau}$, {i.e.} a solution describing a
space with negative constant curvature and with the condition
$p=2$ imposing a restriction on the possibles values of the
cosmological constant. In three dimensions there is a similar
solution \cite{quien} and it corresponds to an extreme
black hole solution. I would like to thank J. Zanelli by
describing these results to me. }
\begin{equation} 
ds^2 = d\tau^2 - {1\over \tau^p} dx_1^2, \label{metrica}
\end{equation}
whereas the general solution for the case $i)$ is regular
everywhere independently of the values of $\alpha $ or $\beta$.
The last point, has an analogue with some general regular
cosmological solutions found in the literature \cite{senovilla}.

Now we quantize the models discussed above; we assume as in
\cite{yo} like-particle boundary conditions and, as a
consequence, there are not anomalies in the functional measure.

The propagation amplitude in the proper-time gauge in the
Euclidean space is 
\begin{equation} 
G [\chi_2, \chi_1; \psi_2, \psi_1] ~=  
\int_0^\infty \prod_x d N(x) \int {\cal D} \psi {\cal D} \chi 
e^{- \int d^2x [{1\over 2N} {({\dot \psi}^2 + {1\over 4}
{{\dot \chi}}^2) + {1\over 8} \mu^2 Ne^{{\alpha \psi + \beta
\chi}}}]}. \label{free}
\end{equation} 

This formula permits to determine the Green function for 
two metric configurations and one can interpret 
(\ref{free}) as describing a set of infinite massless
relativistic particles interacting with an external potential
moving on a two dimensional target space.

In order to compute (\ref{free}) one can use the 
\lq \lq decoupling gauge" $\alpha = 0,\, \beta =1$ and one
obtain  
 
\begin{equation} 
G [\chi_2, \chi_1; \psi_2, \psi_1] ~ 
= \prod_k^\infty \biggl( \int_0^\infty dN_k N_k^{-{1\over 2}} 
e^{-{{(\Delta \psi_k)}^2\over 2N_k}} {\cal G} [ \chi_{2k},
\chi_{1k}; N_k ] \biggr), 
\label{total} 
\end{equation}  
where ${\cal G} [ \chi_{2k},\chi_{1k};N_k ]$ is the propagation 
amplitude for Liouville quantum mechanics, namely 
\begin{equation} 
{\cal G} [ \chi_{2k},\chi_{1k};N_k ] = 
\int {\cal D} \chi_k e^{-  \int d\tau ( {1\over 8N_k}
{\dot \chi_k}^2 + {1\over 8} \mu^2 N_k e^{\chi_k} )}. 
\label{liouville} 
\end{equation}

This formula is formally equivalent to an ordinary quantum 
mechanical problem described by the action 
\begin{equation} 
S_k = \int d \tau \biggl( {1\over 8N_k} {\dot \chi_k}^2 + {1\over 8} \mu^2 
N_k e^{\chi_k}\biggr), 
\end{equation}
with $N^{-1}_k (0)$ playing the role of mass.

The calculation of (\ref{liouville}) is performed 
more easily solving the Schr\"odinger equation 
\begin{equation} 
\biggl[ -{d^2 \over d \chi^2_k} -  \mu^2 e^{\chi_k} 
\biggr] \Psi [\chi_k] =  p_k^2 \Psi [\chi_k] \label{schr}
\end{equation} 
where $p_k$ is the momentum of the $k-th$ particle and 
inserting this solution in 
\begin{equation} 
{\cal G}~[\chi_{2k}, \chi_{1k}; N_k] = 
\int_0^\infty d p_k  e^{-2N_k p^2_k} \Psi^{\ast} (\chi_{2k} ) 
\Psi (\chi_{1k}). \label{G}
\end{equation} 

One can solve (\ref{schr}) distinguishing two cases {\it v.i.z.} 
$i)\, \mu^2< 0$ and $ii) \, \mu^2 >0$. 

When $i)$ is satisfied the solution of the Schr\"odinger equation  is 
\cite{dhoker},\cite{ghandour},\cite{grosche} 
\begin{equation} 
\Psi [\chi_k] = \sqrt{\sinh 2\pi p_k} K_{2ip} (2\sqrt{\mu^2} 
e^{\chi_k/2}), \label{bessel}
\end{equation} 
where $K_{2ip}$ is the modified Bessel function.

Using (\ref{G}) and (\ref{bessel}), (\ref{total}) become  
\begin{eqnarray}  
 G[\chi_2, \chi_1 ; \psi_2, \psi_1] 
&=& \prod_k^\infty \biggl( \int_0^\infty dN_k  
N^{-1/2}_k e^{-{(\Delta \psi_k)}^2/2N_k} \times \nonumber \\
&\times &\int_0^\infty dp_k 
e^{-2{p_k}^2 N_k} K^{\ast}_{2ip} (2\sqrt{\mu^2} e^{\chi_{2k}/2}) 
K_{2ip} (2\sqrt{\mu^2} e^{\chi_{1k}/2}) 
\biggr), \label{liou} 
\end{eqnarray}

The next question is, how one can know if (\ref{liou}) 
is a correct formula?. The answer is easily obtained observing that 
in the limit $\mu^2 \rightarrow 0$, it must reproduce
the propagator for an infinite set of massless
relativistic particles. In such limit, $K_\nu \sim x^{-\nu}$ 
and the expected result is obtained.

In (\ref{liou}), the exponential 
$e^{-{(\Delta \psi)}^2/2N}\rightarrow 0 $ when 
$\tau \rightarrow 0 $ due the solutions of the
equations of motion are singular. The vanishing of this Green 
function say us that the temporal singularity never is 
reached and, in consequence, quantum mechanics smooths the classical
singularities and the big-bang concept could, simply, be 
considered as a classical extrapolation.

It pleasure for me to thank Jorge Zanelli for useful
discussions. This work was partially supported by Grants
FONDECYT (1950278) and DICYT.

\end{document}